\documentclass[
10pt,
twocolumn,          
prl,                         
superscriptaddress
]{revtex4-2}

\usepackage{amsmath,amssymb,bm}
\usepackage{braket}
\usepackage{graphicx}
\usepackage[version=4]{mhchem}
\usepackage{mathrsfs}
\usepackage{xcolor}
\usepackage{soul}

\usepackage[
colorlinks=true,
linkcolor=blue,
citecolor=blue,
urlcolor=blue
]{hyperref}

\begin{document}
\setstcolor{red}	
	\preprint{APS/123-QED}
	
	\title{Chiral switching of elastic spin via dynamic encirclement of exceptional points}
	
	\author{Chenwen Yang}
	\affiliation{Department of Physics, City University of Hong Kong, Kowloon, Hong Kong, China}
	
	\author{Shubo Wang}
	\email{Email: shubwang@cityu.edu.hk}
	\affiliation{Department of Physics, City University of Hong Kong, Kowloon, Hong Kong, China}

\begin{abstract}
Dynamically encircling exceptional points (EPs) enables chiral state conversion in classical wave systems. However, whether this mechanism can be extended to chiral spin conversion has remained elusive. Here we demonstrate chiral switching of elastic spin via dynamic encirclement of EPs in a non-Hermitian micropolar (Cosserat) metamaterial. The interplay between micropolar chirality and anisotropic loss generates EPs with a nontrivial Riemann-sheet topology. Encircling these EPs converts the elastic spin, with the final spin sign dictated solely by the handedness of the encircling trajectory. Our results establish a fundamental route for the selective manipulation of elastic spin, opening avenues for non-Hermitian spin phononics and broader applications in other wave systems.

\end{abstract}
\maketitle

Non-Hermitian systems exhibit unique spectral properties with broad applications in classical wave manipulation~\cite{Making_sense,review_opt_NatPhoto_2017,review_opt_NatPhys_2019,review_opt_CommunPhys_2019,review_Acoustic_Nonhermitian_NRP_2024}. These systems are characterized by complex eigenvalues that form Riemann sheets and coalesce at exceptional points (EPs) \cite{Non-hermitian_chaotic,EP,Obsrv_of_EP,review_EP_Alu_2019}. At these singularities, the system response changes abruptly, giving rise to a variety of unusual physical effects, \cite{review_OptEp_2021}, including unidirectional transmission~\cite{Unidir_PRL_2011,Unidir_NM_2013,Unidir_Acpr_PRM_2018,czzhangOE,Unidir_Acpr_CPL_2023}, topological engineering~\cite{Topo_Nature_2016,Topo_Science_2022,Topo_Ac_PRApp_2024}, and enhanced sensing~\cite{Sensitive_PhotoRe_2020,Sensitive_Ac_MN_2025,Sensitive_Nature_2017,ArbitraryOrder_NC_2019}. Beyond these static properties, the dynamic evolution of states around EPs can produce striking phenomena. In particular, dynamically encircling EPs enables chiral state conversion~\cite{Winding_NC_2018} and asymmetric state transfer~\cite{Dynamic_Encircling_EP_Nature_2016,Dynamic_Encircling_EP_PRL_2017,PhysRevX.8.021066}, with the final state determined solely by the handedness of the encircling trajectory. This mechanism therefore offers a robust approach for asymmetric wave control and has been realized in both optical~\cite{Dynamically_EP_OAM,asymmetric_opt_LSA_2026} and acoustic~\cite{Ac_chiral_switch_APL_2023,Ac_chiral_switch_EML_2024} platforms.

Angular momentum manipulation underpins a variety of fundamental wave phenomena and important applications, such as spin-orbit interactions~\cite{bliokh2015,swang21,ywan2026}, chiral particle sorting~\cite{sbwang2014,zman2024,lai2024observation}, optical/acoustic communications \cite{jwang2012,cshi2017}, and quantum entanglement \cite{emanuel2020}. Recently, dynamical encirclement of EPs has been exploited to control orbital angular momentum, enabling efficient, path-dependent topological charge conversion on chip~\cite{Dynamically_EP_OAM}. However, whether this mechanism can be extended to spin conversion has remained an open question, despite significant efforts in polarization manipulation~\cite{Dynamic_Encircling_EP_PRL_2017,Riemann-Encircling,chirality_enabled_switching}.

Elastic waves have a full vectorial nature with abundant degrees of freedom for exploring spin physics~\cite{IntrinsicElSpin,ElasticAm,FromEs2Ps,Revealing_3DAM_NC_2026}. The manipulation of elastic spin—originating from the circular polarization of the displacement field—is essential for developing spin-phononic devices~\cite{Valley_PRL_2022,Hybrid_PRL_2023,ElS_in_Rod_IJMS_2024,CIPSS_PNAS_2024}. Although non-Hermitian mechanisms have been extensively leveraged to manipulate elastic wave propagation~\cite{NonHermitian_Odd_PRR_2020,NonHermitian_elasto_APL_2024,Exp_TPModulation_PRL_2019,Exp_TPModulation_PRApp_2020,NonHermitian_wavedynmc_JMPS_2024,cao2025asymmetric,ChiralModeConversion_JMPS_2026}, elastic spin conversion via encircling EPs has remained elusive due to the fundamental constraint enforced by the symmetry of the stress tensor, which forbids the coupling of orthogonal linear polarizations. A promising route to overcoming this restriction is the use of micropolar (Cosserat) elasticity~\cite{Noncentrosymmetry_IJES_1982}. Unlike conventional materials, micropolar media naturally incorporate rotational degrees of freedom alongside translation~\cite{3DMicropolar_Science_2017,Micropolar_HGK_2012}. This unique feature endows them with an inherent structural chirality~\cite{Micropolar_theory_2017,Micropolar_chiral_JMPS_2018,Micropolar_chiral_NC_2019,qtong23}, providing a natural platform for engineering the non-Hermitian Hamiltonians required for spin manipulation.

In this work, we demonstrate selective chiral switching of elastic spin by employing a micropolar medium with intrinsic chirality and anisotropic loss.  We analytically derive the effective Hamiltonian that captures the coupling between orthogonal linear polarizations in the medium and identify the EPs in parameter space. We show that dynamically encircling the EPs enables chiral conversion of elastic spin, with the final spin sign solely dictated by the handedness (e.g., clockwise or counterclockwise) of the encircling trajectory. These theoretical predictions are validated through full-wave finite-element simulations of elastic spin transport in a realistic micropolar metamaterial. Our work establishes a convenient platform for asymmetric manipulation of elastic spin, opening a route toward robust mechanical wave routing and chiral phononic applications.

\textcolor{black}{\textit{Model system}---}We consider the elastic waves supported by an elastic rod extending along the $x$-axis with mass density $\rho$, cross-section area $A$, Young's modulus $E$, and moment of inertia along the $i$-axis $I_i$. According to the Euler-Bernoulli beam theory, the dynamical equation governing its displacement in the transverse direction, $u_y$ and $u_z$, can be written as~\cite{SI}:
\begin{align}
	\rho A\partial_t^2 u_y + EI_z\partial_x^4 u_y+ \gamma_y\partial_t u_y+ c\partial_x^3 u_z &= 0,\nonumber\\
	\rho A\partial_t^2 u_z + EI_y\partial_x^4 u_z + \gamma_z\partial_t u_z - c\partial_z^3 u_y &= 0.
\label{eq_dynamic}
\end{align}
Here, $\gamma_y$ and $\gamma_z$ are the damping coefficients, and $c$ is the chiral coefficient. The first two terms of the above dynamical equations correspond to the standard form of the Euler–Bernoulli beam theory under free vibration, and the remaining two terms represent the contributions from damping and chirality, respectively. 

\begin{figure}
	\includegraphics[width=1\linewidth]{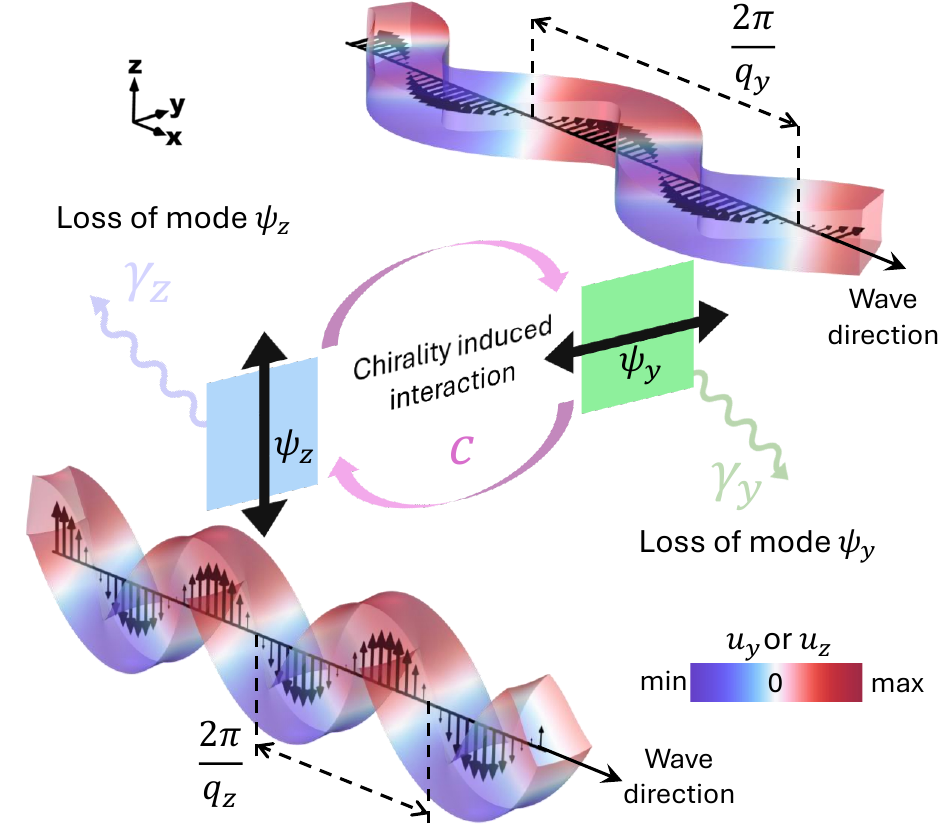}
	\caption{ 
		Schematic of the theoretical model and key parameters. An elastic beam aligned along the $x$-axis supports two transverse vibrational degrees of freedom, $u_y$ and $u_z$, indicated by the solid black arrows. The corresponding transverse modes are $\psi_y$ (green plane) and $\psi_z$ (blue plane), respectively. In the absence of intrinsic losses ($\gamma_y$, $\gamma_z$) and chirality-induced coupling ($c$), their respective wavenumbers are $q_y$ and $q_z$. These wavenumbers are governed by the cross-sectional moments of inertia about the $z$- and $y$-axes, respectively, and act as effective on-site potentials in the system.
	}
	\label{system}
\end{figure}

To gain an intuitive understanding of the model system, we apply the slowly varying envelope approximation under the assumption that the damping and chirality are small. The time-harmonic solutions to the above equations take the form of $u_m(x,t) = \psi_m(x) e^{-i\omega t}$, where $m=y,z$ and $\omega$ is the angular frequency. In the absence of coupling and damping, the intrinsic wavenumbers for the two polarization directions are: $q_y = \left( \rho A \omega^2/(E I_z) \right)^{1/4}, q_z = \left( \rho A \omega^2/(E I_y)\right)^{1/4}.$ The dynamic equation can be written as \(\partial_x \ket{\psi} = i H \ket{\psi}\) with the effective Hamiltonian~\cite{SI}
\begin{equation}
H= \begin{pmatrix} 
		q_y + i \Gamma_y & i \chi_y \\ 
		-i \chi_z & q_z + i \Gamma_z 
	\end{pmatrix}, 
\label{Schrodinger}
\end{equation}
where $\ket{\psi}=(\psi_y,\psi_z)^T$; $\Gamma_y=\omega\gamma_y/(4EI_z q_y^3)$; $\Gamma_z=\omega\gamma_z/(4EI_y q_z^3)$;
$\chi_y=c q_z^3/(4EI_z q_y^3)$; $\chi_z=c q_y^3/(4EI_y q_z^3)$.
At this point, we can see that $q_m$, $\gamma_m$, and $c$ mainly influence the real part of on-site potential, the damping, and the coupling between $\psi_y$ and $\psi_z$ (as illustrated in Fig.~\ref{system}), respectively. Equation~(\ref{Schrodinger}) indicates that an EP emerges when $q_y=q_z$ and $\left(\Gamma_y-\Gamma_z\right)^2=4\chi_y \chi_z$.

\begin{figure}
	\includegraphics[width=1\linewidth]{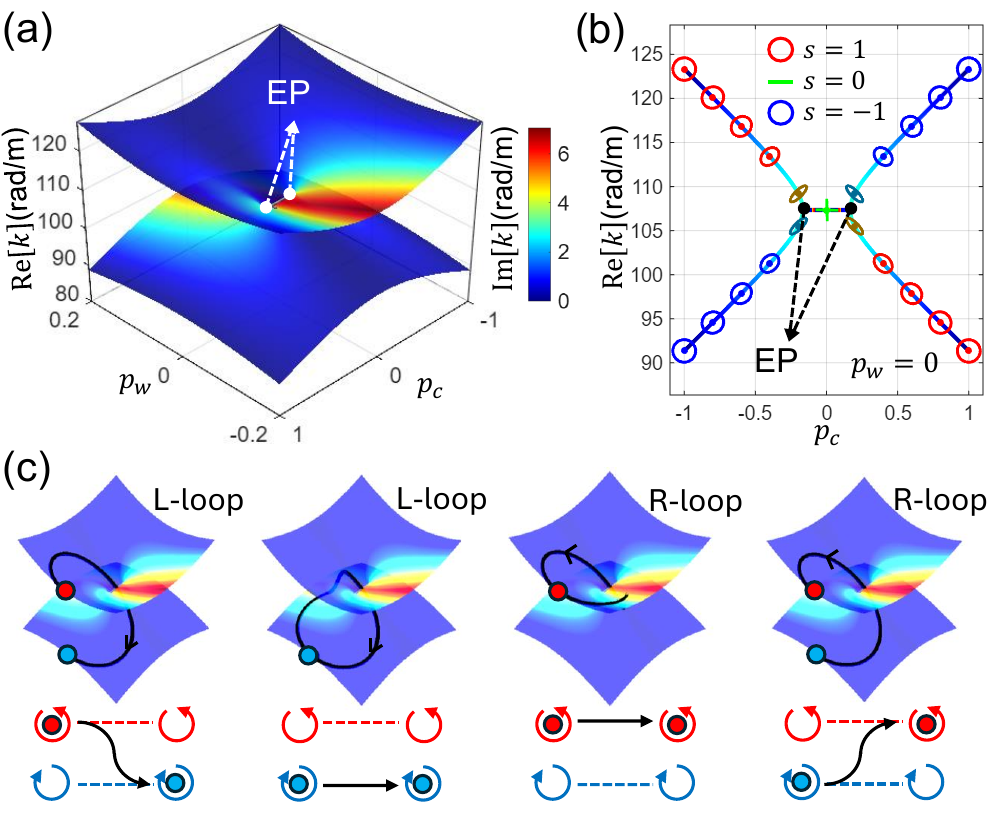}
	\caption{ {{Chiral switching of elastic spin.}} 
		(a) Eigenvalues (i.e., wavenumbers) as functions of the anisotropy parameter $p_w$ and chirality paramater $p_c$  (b) Eigenstates at $p_w=0$. The closed curves denote the polarization ellipses of the eigenstates in the $y$–$z$ plane. The color of the eigenvalue lines represents the magnitude of ${\rm Im}[k]$. (c) Schematic for the chiral switching of elastic spin. Red (blue) dot denotes eigenstate with $s=+1$ ($s=-1$). The black solid lines on the Riemann sheets indicate the evolution trajectories, with arrows marking the evolution direction. All four trajectories start from $(p_w,p_c)=(0,1)$, pass through $(0,0)$, and return to $(0,1)$.
	}
	\label{theory}
\end{figure}


\begin{figure*}
	\includegraphics[width=1\linewidth]{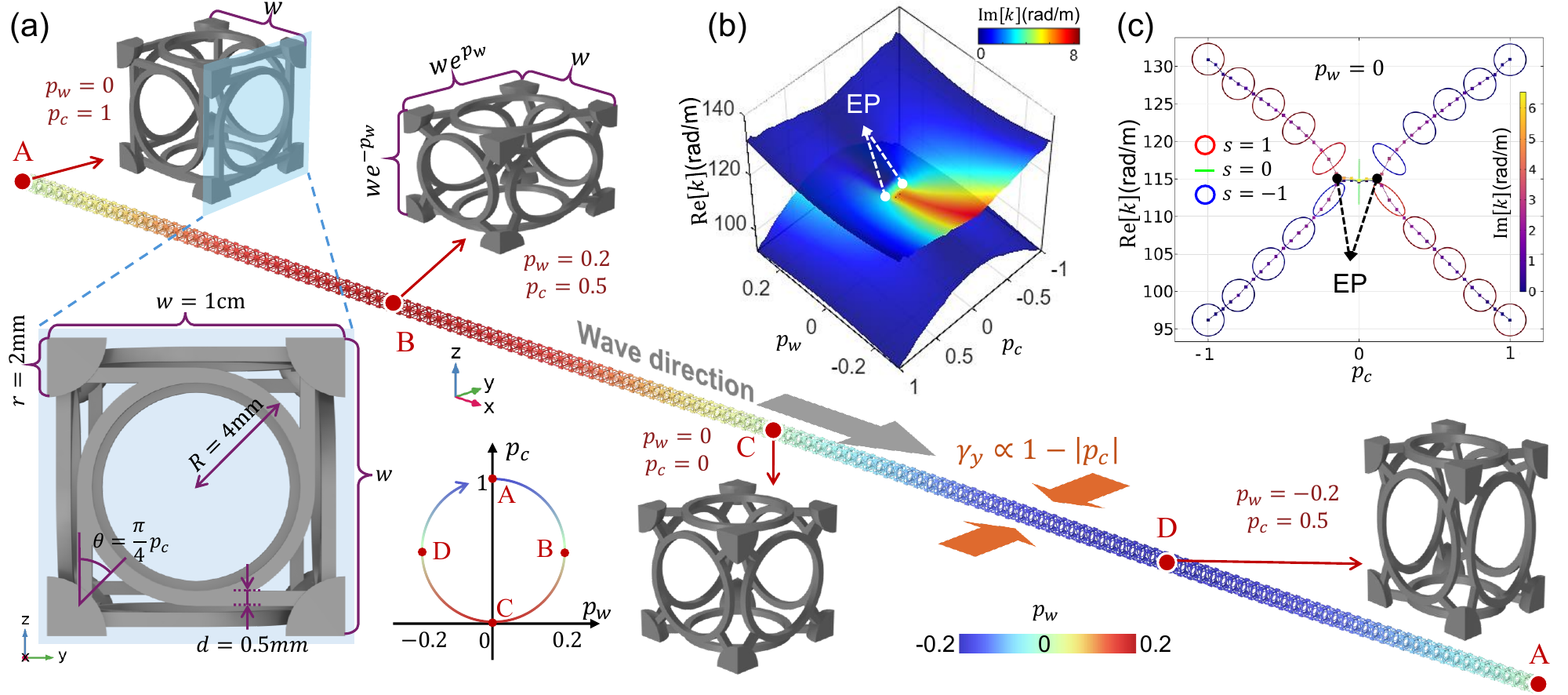}
		\caption{ Non-Hermitian micropolar metamaterial beam.
		(a) Illustration of geometric parameters and setups. The metamaterial beam consists of 200 micropolar unit cells arranged along the $x$ direction. Each hexahedral unit is formed by octant-ball corner blocks (radius $r$) connected by rings and linking beams. The rings have outer radius $R$, and the linking beams have thickness $d$. The chirality of each unit is controlled by $\theta$, with $\theta=\pi/4,\,0,\,-\pi/4$ corresponding to $p_c=1,\,0,\,-1$, respectively. Additional scaling along the $y$ and $z$ directions is controlled by parameter $p_w$. Damping is applied on the outer flat surfaces of the corner blocks. The material is aluminum, the operating frequency is $3000\,{\rm Hz}$, and the applied maximum damping coefficient along $y$-axis is  $10\,{\rm kN\cdot s/m^3}$. Dynamic state evolution is induced by varying the parameters as $p_w=0.2\sin(2\pi x/L)$ and $p_c=0.5+0.5\cos(2\pi x/L)$ for $x\in[0,L]$ and $L=200 w$. Representative unit cells at positions A–D are displayed in the insets. (b) Simulated eigenvalue spectra. (c) Polarization ellipses of the averaged displacement field $\bf u$ of the eigenstates in the $y$–$z$ plane.
	}
	\label{setup}
\end{figure*}


We now show how to construct a parameter space within which the encircling of EPs gives rise to the chiral switching of elastic spin. For the emergence of an EP, the system must incorporate a non-zero anisotropic loss (i.e., $\gamma_y \neq \gamma_z$), and the loss anisotropy should balance the micropolar chirality (i.e., $\left(\Gamma_y-\Gamma_z\right)^2=4\chi_y \chi_z$). Small loss and weak chirality should be adopted near the EP to avoid strong nonadiabatic effects during dynamic evolution~\cite{WithOutEP}. Meanwhile, the initial and final states of the dynamic evolution should be circularly polarized eigenstates $\ket{\psi}=(1,\pm i)^T$ carrying opposite normalized spin $s=\left({\rm Im}[{\bf u}^*\times{\bf u}]/|{\bf u}|^2\right)_x=\pm1$~\cite{IntrinsicElSpin,ElasticAm}, which requires $c \neq 0$, $q_y=q_z$, and $\gamma_{y,z}=0$. Considering the above requirements, we introduce two dimensionless parameters, $p_w$ and $p_c$, to control the material anisotropy and chirality, respectively. Specifically, we set $q_y/q_z = e^{-p_w}$, $q_y q_z=q_0^2$, $c=c_{\max}p_c$, and $\gamma_y=\gamma_{\max}(1-|p_c|)$ with $\gamma_z=0$. Here, $q_0$, $c_{\max}$ and $\gamma_{\max}$ are constants. As such, the parametric space $(p_w, p_c)$ hosts a pair of EPs near $(0,0)$ while ensuring that the initial and final states at $(0,1)$ are circularly polarized states with $s=\pm 1$. We note that $c_{\max}$ should be large enough to guarantee a sufficiently large eigenvalue splitting between $s=+1$ and $s=-1$ states for robust chiral conversion. This does not compromise the validity of the effective Hamiltonian, which assumes a small $c$ near the EPs, and nor does this alter the essential topology of the eigenvalue spectra.

Figure~\ref{theory}(a) shows the eigenvalue (i.e., wavenumber) spectra corresponding to the Hamiltonian in Eq. (2). We notice that two EPs emerge symmetrically at $p_w=0$ with a typical Riemann-sheet topology. \textcolor{black}{Figure~\ref{theory}(b) shows the eigenstate polarization for $p_w=0$, where the states are predominantly circular across the symmetry-preserving phase and predominantly linear across the symmetry-breaking phase}.

Based on the Riemann-sheet spectral topology, the dynamic evolution of eigenstates tracing a closed circle around either EP in the parameter space can enable chiral switching of elastic spin. We note that the chiral switching refers to the handedness of the evolution trajectory rather than the micropolar chirality. Taking the left EP as an example, a clockwise (left-handed, denoted as ``L-loop") or anticlockwise (right-handed, denoted as ``R-loop") evolution of the same initial state, starting from $(p_w,p_c)=(0,1)$ and passing through $(0,0)$, will result in different final states. Specifically, as illustrated in Fig~\ref{theory}(c), the initial state with $s=+1$ (red dot) evolves into the $s=-1$ ($s=+1$) state along the L-loop (R-loop). Similarly, the initial state with $s=-1$ (blue dot) also evolves into the $s=-1$ ($s=+1$) state along the L-loop (R-loop). Therefore, the sign of the final spin only depends on the handedness of the encirling paths. This property is attributed to the Riemann-sheet topology and the loss-induced nonadiabatic transition near the branch point, which favors transition from the higher-loss sheet (red area) to the lower-loss sheet (blue area).


\begin{figure*}
	\includegraphics[width=1\linewidth]{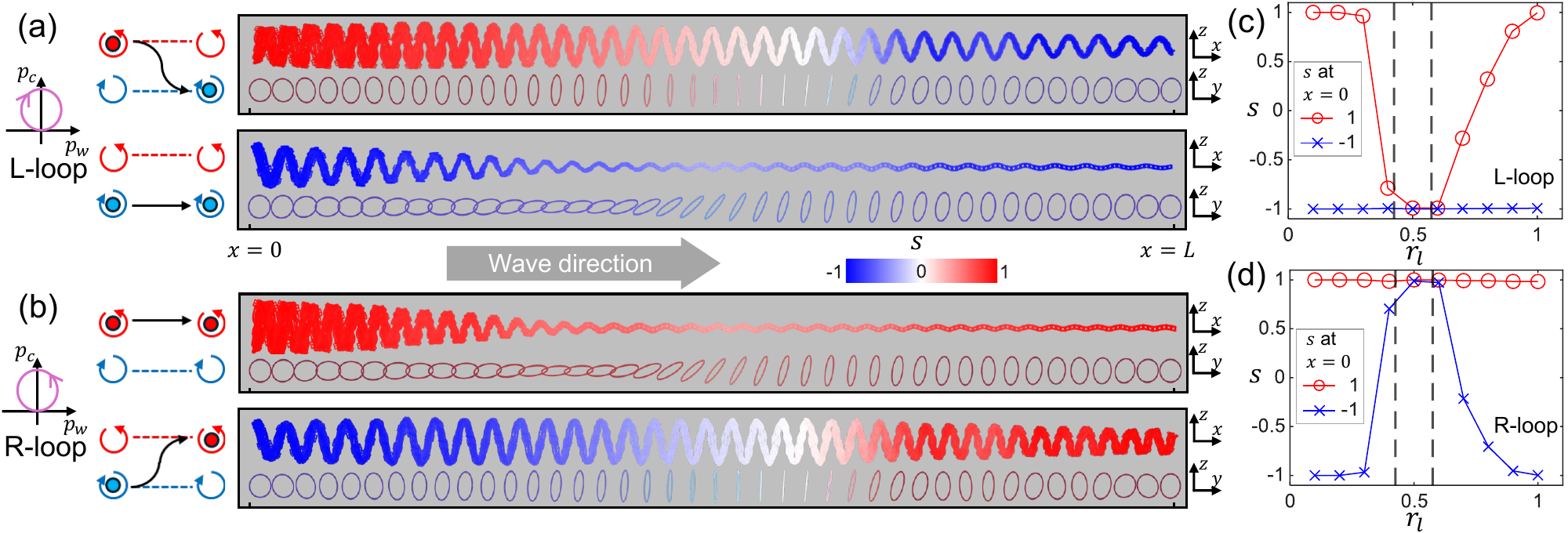}
	\caption{ {Chiral switching of elastic spin in the micropolar metamaterial.} (a) L-loop and (b) R-loop encirclings of the EP, starting from $(0,1)$ and passing through $(0,0)$ in the parameter space, with propagation from left to right (along $+x$-axis). In each inset (gray background), the upper panel shows the beam deformation, colored by $s$; the lower panel shows the corresponding polarization in $y-z$ plane. For the L-loop, $s=+1$ evolves into $s=-1$, while $s=-1$ remains unchanged. The R-loop shows the opposite behavior, selectively converting $s=-1$ to $s=+1$. (c) and (d) show the spin $s$ at $x=L$ for different $r_l$ with L-loop and R-loop, respectively. The black dashed lines mark the $r_l$ where the loops intersect the EPs.
	}
	\label{results}
\end{figure*}

\textcolor{black}{\textit{Realization with a micropolar metamaterial}---}The theoretical model can be implemented with the micropolar metamaterial beam in Fig.~\ref{setup}(a). The metamaterial has microstructures with a chirality continuously tunable via geometric parameters~\cite{Micropolar_theory_2017,Micropolar_chiral_NC_2019}. Specifically, since $q_{y,z} \propto 1/I_{z,y}$, the real part of the on-site potential can be tuned via the ratio of the moment of inertia $I_y/I_z$. As such, the side lengths of the unit cell along the $y$- and $z$-directions can be set as $w e^{p_w}$ and $w e^{-p_w}$, respectively, yielding $q_y/q_z = e^{-p_w}$. Independently tunable velocity-proportional damping can be realized via eddy-current effects in conductors under magnetic fields~\cite{Damping_theory1,Damping_theory1}, which can generate strong anisotropic damping~\cite{Damping_exp1,Damping_exp2}. For simplicity, we directly introduce the damping coefficients $\gamma_y$ and $\gamma_z$ into the metamaterial in numerical simulations. The micropolar chirality $c$ is controlled by the angle $\theta=\pi p_c/4$ through the parameter $p_c$. Along the wave propagation direction, the geometric parameters $(p_w,p_c)$ gradually change from $(0,1)\rightarrow(0.2,0.5)\rightarrow(0,0)\rightarrow(-0.2,0.5)\rightarrow(0,1)$, following a circular trajectory as shown in the bottom inset of  Fig.~\ref{setup}(a). This realizes the encircling of the EP along the L-loop.

We perform full-wave numerical simulations of the metamaterial unit cell using the finite-element-method package COMSOL, assuming periodic boundary condition along the $x-$axis. The obtained eigenvalue spectra are shown in Fig.~\ref{setup}(b) for different parameters $(p_w, p_c)$, which agree well with the theoretical predictions in Fig.~\ref{theory}(a). For the eigenstates' polarization, we take the averaged displacement field over a plane perpendicular to the $x$-direction within a unit cell and calculate its polarization ellipse. The results for the eigenstates on the $p_w=0$ plane are shown in Fig.~\ref{setup}(c), which also exhibit good agreement with the theoretical results. These results demonstrate the validity of the micropolar metamaterial in realizing the model.

We further simulate the dynamic encircling of EPs for chiral switching of elastic spin, using the metamaterial beam (total length $L$) in Fig.~\ref{setup}(a). Absorbing boundary conditions are imposed at both ends of the beam. The initial state at the left end ``A" is a circularly polarized elastic wave. The wave propagates from $x=0$ to $x=L$, which is mapped onto an eigenmode evolution trajectory in the parameter space. Figures~\ref{results}(a,b) show the results for the L-loop and R-loop defined by $p_w=0.2\sin(2\pi x/L)$ and $p_c=0.5+0.5\cos(2\pi x/L)$ ($h=\pm1$ for L/R-loop), which cross the origin $(p_w,p_c)=(0,0)$. In both cases, we plot the beam deformation  (upper panel) and polarization ellipse of the averaged displacement field (lower panel). Clear selective chiral switching of elastic spin is observed. In the case of L-loop, an initial state $s=+1$ gradually evolves into the state $s=-1$, while an initial state $s=-1$ remains in the same state. In contrast, in the case of R-loop, an initial state $s=-1$ gradually evolves into the state $s=+1$, while an initial state $s=+1$ remains in the same state. Additionally, we notice that the deformation amplitude gradually reduces while the wave propagates in all the cases due to dissipation.

We then investigate the influence of the radius $r_l$ of the circular loops. The evolution trajectories are parameterized as $p_w=r_l\,h\,0.4\sin(2\pi x/L)$ and $p_c=1-r_l+r_l\cos(2\pi x/L)$. When $r_l=0.5$, the path passes through $(p_w,p_c)=(0,0)$. Figures~\ref{results}(c,d) show the final states  as functions of the loop radius $r_l$. The black dashed lines mark the $r_l$ values at which the loop intersects the two EPs. The results show that selective chiral switching is most pronounced when the evolution trajectory traverses the branching region of the Riemann sheets between the two EPs. In addition, the results indicate that partial spin switch (with the final spin between $s=-1$ and $s=+1$) occurs even when the trajectories pass near the EPs without strictly encircling them. This results from the non-adiabatic effects due to the finite beam length, as well as the fact that such chiral switching relies more on the local Riemann-sheet structure than on the strict topological winding~\cite{WithOutEP}.

\textit{Conclusion---}We demonstrate the selective chiral switching of elastic spin in a non-Hermitian micropolar metamaterial. By introducing anisotropic moduli and material loss into an intrinsically chiral micropolar medium, we construct the Riemann sheets and exceptional points (EPs) required for asymmetric spin conversion. Our work establishes non-Hermitian micropolar metamaterial as a versatile platform for elastic spin manipulation. The ability to deterministically control spin degrees of freedom opens new avenues for chiral mechanical signal processing. These results can be readily applied to the design of spin-phononic devices, such as spin-dependent elastic-wave routers, phononic spin filters, and robust logic gates. The underlying principles may be generalized to other wave systems, such as optical systems.

~\\
\textit{Acknowledgments}---The work described in this paper was supported by grants from the Research Grants Council of the Hong Kong Special Administrative Region, China (Project No. AoE/P-502/20) and National Natural Science Foundation of China (No. 12322416). 

~\\
\textit{Data availability}---The data that support the findings of this study are available from the corresponding authors upon reasonable request.

\bibliography{ref}

\end{document}